\documentclass[letterpaper,twocolumn,10pt]{article}
\usepackage{fullpage}
\usepackage{usenix}
\usepackage{times}
\usepackage{authblk}
\usepackage{hyperref}
\hypersetup{pdfstartview=FitH,pdfpagelayout=SinglePage}

\usepackage{xcolor}%
\usepackage{url, subcaption}
\usepackage{comment}

\newcommand{\ddc}[1]{\textcolor{brown}{\emph{#1 --DDC}}}
\newcommand{\kc}[1]{\textcolor{violet}{\emph{#1 --kc}}}
\newcommand{\mjl}[1]{\textcolor{brown}{\emph{#1 --mjl}}}
\newcommand{\todo}[1]{\textcolor{red}{\emph{#1}}}

\usepackage{graphicx}
\usepackage{booktabs}
\usepackage{soul}
\usepackage[title]{appendix}

\providecommand{\ver}{VERIFIED}
\providecommand{\bgpsec}{BGPsec}

\begin{document}
\hyphenation{vulner-abil-ity}
\date{}

\title{A path forward: Improving Internet routing security by enabling zones of trust}

\author[1]{David Clark}
\author[2]{Cecilia Testart}
\author[3]{Matthew Luckie}
\author[3]{KC Claffy}
\affil[1]{MIT / CSAIL}
\affil[2]{Georgia Tech}
\affil[3]{UC San Diego / CAIDA}
\maketitle

\begin{abstract}

Although Internet routing security best practices
have recently seen auspicious increases in uptake, 
ISPs have limited incentives to deploy them. 
They are operationally complex and expensive to
implement, provide little competitive advantage, and protect only
against origin hijacks, leaving unresolved the more general threat of
path hijacks.  We propose a new approach that achieves
four design goals: improved incentive alignment to implement
best practices; protection against
path hijacks; expanded scope of such protection to 
customers of those engaged in the practices; and reliance
on existing capabilities rather than needing
complex new software in every participating router. 

Our proposal leverages an existing coherent core of interconnected
ISPs to create a {\em zone of trust}, a topological region that protects
not only all networks in the region, but {\em all directly
attached customers} of those networks.  Customers benefit from
choosing ISPs committed to the practices, and ISPs thus
benefit from committing to the practices.  We compare our
approach to other schemes, and discuss how a related proposal,
ASPA, could be used to increase the scope of protection our
scheme achieves.  We hope this proposal inspires discussion
of how the industry can make practical, measurable progress
against the threat of route hijacks in the short term by
leveraging institutionalized cooperation 
rooted in transparency and accountability.

\end{abstract}

\section{Introduction}
\label{sec:intro}

The Internet's global routing protocol -- Border Gateway  Protocol (BGP) --
suffers from a well-documented vulnerability: a network (termed an
Autonomous System or AS) can falsely announce that it hosts
or is on the path to a block of addresses that it does not in fact have
the authority to announce.
Routers that accept a forged route announcement -- known as a
\emph{route hijack} -- will route traffic intended for
addresses in that block to a rogue AS.
The simplest form of route hijack is an {\em
origin hijack}, in which a malicious AS falsely announces (`originates
an assertion') that it directly hosts (i.e., is the origin for) a prefix
that belongs to someone else.  In a {\em path hijack}, an attacker claims to be an AS {\em in} the path to a prefix, forging the legitimate owner's ASN as the origin of the prefix.
The highly distributed operation of the BGP protocol -- $\approx$75K
independent networks around the world -- and its role in establishing
and maintaining the connectivity we call ``the Internet'', have
contributed to the persistence of this long-standing but increasingly
dangerous vulnerability.

The two clear victims of a route hijack are the owner 
of the hijacked block and the sender of traffic to the hijacked block.  
If the attacker hijacks address space in order to impersonate the legitimate
holder~\cite{apostolaki_hijacking_2017,goodin_suspicious_2018,ars-hijacks-2022,
ripe_network_coordination_centre_youtube_2008}
or to inspect~\cite{paganini_hijacking_2017} the traffic, then 
senders of traffic to the hijacked block may fall victim to a scam or 
surveillance.
If the attacker hijacks address space in order to conduct malicious
activity~\cite{ramachandran_understanding_2006,vervier_mind_2015,white_ops_hunt_2018},
a third victim is the target of the malicious activity.  
The malicious activity may cause blocklisting of the address block,
which impairs the legitimate owner's use of the block.  

The best currently available practices in routing security require 
two steps to identify and block propagation of
bogus route announcements.  First, each ISP must register its own
address space in a trusted database (ideally, the Resource Public 
Key Infrastructure aka RPKI) and routers across 
the Internet must check announcements against
such a database and drop those announcements that do not match
({\em route origin validation} aka ROV). 
An AS who engages in the first step gains no security unless
other ASes correctly deploy the second (ROV) step. 

ROV is sufficiently
operationally complex that smaller or lower-resourced ISPs are reluctant
to risk misconfigurations that impair their own service availability.
Thus, networks take on additional costs and operational
risks, but the benefits may not accrue to them or their customers.  
Even if consistently implemented, which is a lofty aspiration in a global
context, the practices target only the simplest form of hijack,
an {\em origin hijack}.  The proposed approach for protection against
a broader range of hijacks is \bgpsec, an even more complex and
expensive protocol-based solution that is at least a decade
away from significant operational deployment \cite{bgpsec.net}.

Concerns over slow progress on routing security solutions
led the U.S. Federal Communications Commission (FCC) to issue 
a February 2022 Notice of Inquiry into potential regulatory 
interventions that could reduce the severity of the threat to U.S. 
networks and traffic~\cite{fcc-noi-2022}.
Several U.S. government agencies, including the DHS and
a joint filing by the DOD and DOJ, urged the FCC to take action
\cite{dhs-bgpnoi-2022,doj-bgpnoi-2022}.  Other commenters emphasized the 
risks of regulation in this domain.

Tension is increasing on this topic, as multistakeholder efforts 
to advance routing security have continued 
for over a decade. In the meantime, the risk and prevalence of
both accidental and malicious BGP hijacks grows, rendering even the
largest companies in the world victims of hijacks \cite{ars-hijacks-2022}.
The scope of the problem is elusive to measure given 
lack of disclosure -- and sometimes lack of awareness of -- incidents.

The collective-action characteristic of the problem is fundamental:
even those who are willing to invest in order to increase their own 
routing security cannot achieve protection without commitments from other
networks to prevent propagation of bogus routes.
We propose a more practical solution that refocuses on a new
goal: to {\em provide a concrete action that a security-aware AS
can take to protect itself from both having its address blocks
hijacked, and its traffic to other address blocks hijacked}.

We propose an approach that achieves four related goals.  
First, it aligns incentives of actors toward
improved routing security. 
Second, it offers protection against not only origin
hijacks but the larger looming problem of path hijacks.
Third, it allows ASes participating in
the approach to protect their customers without
additional work on the part of the customer, thus allowing
highly-resourced ISPs to protect other parts of the Internet.  
This feature is compelling because in today's Internet the
steps necessary to securely configure systems 
are sometimes complicated, and smaller ASes may not
have the skills or resources to undertake them.
Even more compelling is the resulting alignment of incentives:
customers will prefer a participating provider since they
offer enhanced security, giving providers an incentive to 
participate so they can market their improved security to 
potential customers. 
Finally, our approach requires no new capabilities
in routers, relying on existing capabilities and institutions,
and current techniques for analyzing interdomain (BGP) topology data.

The roadmap of this paper is as follows.  We first describe 
barriers to routing security
over the last two decades (\S\ref{sec:background}). We describe
the threat model in \S\ref{sec:threat}. In \S\ref{sec:design} 
we introduce the principles of a {\em zone
of trust}, a connected region of the Internet where providers take
enhanced steps to improve the security of that region, including the
security of customers connected to providers in the region.
We introduce a specific example of a routing zone of trust 
which offers a more incentive-aligned direction for 
protecting ASes from both \emph{origin} hijacks and \emph{path} hijacks.

We analyze the residual risks of our scheme and 
how to minimize them (\S\ref{sec:considerations}),
auditing requirements (\S\ref{sec:auditing}), and 
comparison to other proposals (\S\ref{sec:comparison}).

\section{Background and Related Work}
\label{sec:background}

\begin{figure*}
\centering \includegraphics[width=.9\textwidth,height=.30\textwidth]{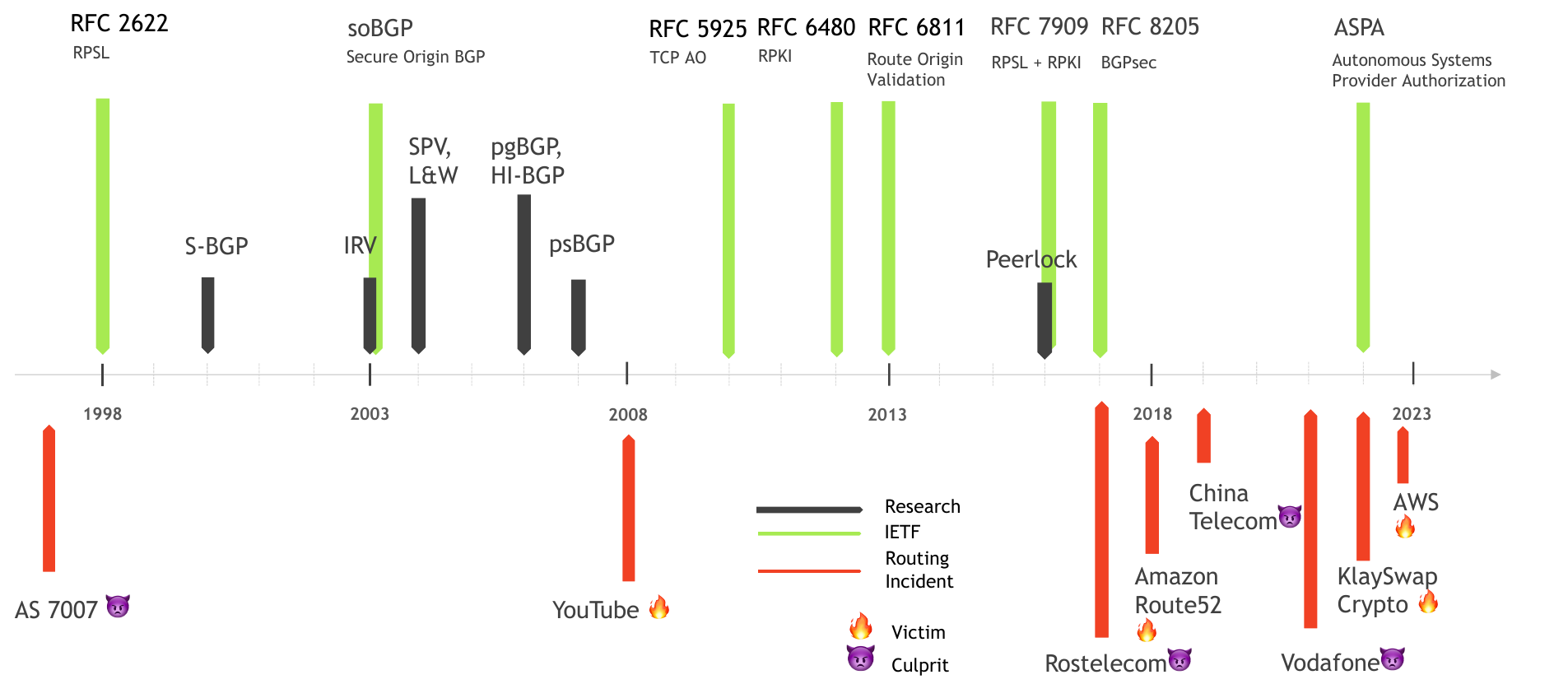}
\caption{Decades of proposed routing security approaches;
	sample of high-profile hijacks.}
\label{fig:background:timeline}
\vskip -5mm
\end{figure*}

The Internet standards community has long struggled with proposals to
tighten the integrity of BGP communications.
As with protection of other Internet transport mechanisms (e.g.,
DNSSEC, TLS), the standards community has grappled
with complexities of cryptographic key management, trust anchors, and
performance implications that hinder standardization,
implementation, and deployment.
Over the last 30 years, over 20 proposals to secure BGP have come 
from academia, industry and the Internet Engineering Task Force
(IETF), some of which Figure~\ref{fig:background:timeline} highlights. 
We describe how the standards and operational communities
have tried to tackle this problem, and how it motivates our proposal. 

\subsection{Interdomain Routing}
\label{sec:background:bgp}

ASes use BGP to exchange {\em routes} that describe paths to
destinations in the global Internet.
Two important components of a route are the {\em prefix} that
specifies the block of addresses of a route, and the
{\em AS path} that reports the sequence of ASes that received
the route.
To prevent forwarding loops, a router chooses the
most specific route to a destination IP address -- i.e.,
for 192.0.31.8, it would prefer a route with a prefix 192.0.31.0/24
over 192.0.30.0/23.
Operators use this property for traffic engineering. 
BGP also provides a mechanism to annotate announcements with attributes
-- known as {\em BGP communities}~\cite{rfc1997} -- to enable
signaling within and across ASes, facilitating
traffic engineering innovations~\cite{BGP-Communities:CCR2008}
such as automated blocking of denial-of-service attack traffic 
on the path to the
victim~\cite{rfc7999,Cisco-remotely-triggered-black-hole-filtering}.

There are two general types of relationship between neighboring
ASes: customer-to-provider (c2p), where the customer pays a provider
to obtain global reachability, and peer-to-peer (p2p), where two peers
exchange routes to their customers without involving an intermediate
provider~\cite{gao01asrel}.
If an AS has multiple routes to the same prefix, the rational choice
is to prefer routes
received from customers (a source of revenue), over routes
received from peers (typically settlement-free, i.e., no cost), 
over routes received from providers (which cost the AS)~\cite{gao01asrel}.
Other ASes that an AS X can reach through a customer link are within the
{\em customer cone} of X. 

A few ($\approx$15) ASes obtain global routing using routes
received only from their peers and customers, i.e., they do
not pay any transit providers.  These ASes connect in a full
mesh (a peering clique) that enables packet delivery between
arbitrary networks with different transit providers.  The ASes
in this group that also do not pay for peering are known as
Tier-1 providers. However, payments between ASes are confidential;
we thus use the term {\em Tier-1} to refer to the peering clique.

\subsection{Routing Security in the 1980s}
\label{subsec:routingsecurity}

In 1982, Rosen~\cite{rosenrfc827} documented that it is possible to
corrupt interdomain routing in RFC 827, in the context of
a predecessor of BGP called the Exterior Gateway Protocol (EGP):
\begin{quote}
\textit{If any gateway sends an NR [neighbor reachability] message with
false  information, claiming  to be an appropriate first hop to a network
which it in fact cannot even reach, traffic destined  to  that  network
may never be delivered.  Implementers must bear this in mind.}
\end{quote}
This warning to implementers suggests the perceived threat in 1982 was
accidental misconfiguration, rather than malicious operators.

\subsection{Routing Security in the 1990s}
\label{sec:background:irr}

To mitigate the prevalent risk of accidental misconfigurations,  
in the 1990s network operators developed the 
{\em Internet Routing Registry (IRR)} system of databases.
The IRR system enabled network operators to publish address
ownership and routing policy information~\cite{IRR-db},
which other operators could use to build filters that permit or
deny routes according to these operator-registered policies.
Unfortunately, some IRR systems do not authenticate registration data, 
allowing attackers to compromise the IRR by falsely registering
ownership of resources which they then use in a hijack~\cite{radbabuseguilmette,radbrevokedip,toonk14spammers,2022-du-ihre,oliver-imc2022}.

When the IETF started to study malicious BGP security threats,
in the late 1990s, they did not initially assume that an AS
operator was an important threat actor.  Instead, they focused
on the threat that a third party could intercept the traffic
between two well-behaved ASes and then modify the BGP update
to inject a false assertion. To defend against this threat,
in 1998 the IETF added an optional extension to TCP to allow
end-points to authenticate the contents of a TCP
segment\cite{TCPMD5,TCPAO}.

\subsection{Routing Security in the 2000s: BGPsec}
\label{sec:background:bgpsec}

Aiming for a more complete approach to routing security, in 2006 the IETF's
Secure Inter-Domain Routing (SIDR) Working Group began designing a variant
of BGP that would support {\em path validation} -- ensuring that each AS 
appearing in a received AS path was legitimately in the path.   
During this decade over a dozen
competing approaches came out of academia and industry~\cite{smith_securing_1996, S-BGP_kent-2000, HopIntegrity-2002, MOAS_List-2002,
soBGP-2003, IRV-2003, SPV-2004, subramanian_listenWhisper-2004,psBGP_2005, 
CCR-bewareBGPattacks-2004, karlin_pretty_2006, ESMs-2006, Hi-BGP-2006, israr_credible_2009}.
The protocol that became an IETF standard (RFC 8205) in 2017 is 
called BGPsec~\cite{bgpsec}. 
BGPsec update messages include two important new fields: 
the AS to which the router is sending that
announcement; and a cryptographic signature over the message that enables
any router along the path to verify that the series of signatures are
valid.  This mechanism prevents path hijacks:  a malicious AS cannot
forge the AS path because the malicious AS cannot sign records 
for the forged ASes.

Cryptographic attestation of paths requires propagation of a 
new layer of cryptographic
transaction at each hop, which is computationally expensive and
poses a router-level (rather than AS-level or prefix-level)
key distribution challenge, since every router must have its
own public key signed by a certificate authority.  Furthermore,
full protection of the path requires 
every AS along the path to implement BGPsec. 
Partial deployment, inevitable during
a transition, implies unpredictable protection. 
The complexity, overhead, and misaligned incentives
have prevented significant operational deployment of BGPsec,
despite a decade-long standardization process that completed
7 years ago \cite{bgpsec}.

\subsection{Routing Security in the 2010s} 

The 2010s brought three areas of endeavor: rigorous analyses of the
incentives to deploy routing security solutions; technology,
standardization, and operational mechanisms to mitigate the simpler
problem of origin hijacks; and a collective action effort 
(MANRS) to overcome
the counter-incentives to deploying these mechanisms.

\subsubsection{Analyzing deployment incentives}
\label{sec:background:incentives}

As early as 2009 researchers began to survey the array of efforts and
analyze why they had failed to gain traction \cite{nicholes_survey_2009,
butler_survey_2010}. Such reviews continued throughout
the subsequent decade~\cite{siddiqui_survey_2015, mitseva_state_2018,testart_reviewing_2018}.
Researchers also explored approaches to overcome the economic 
counter-incentives to deployment of protocol-based approaches to routing
security, and analyzed the implications of partial deployment
\cite{gill2011-market,lychev_bgp_2013,CACM-SecureInternet,cohen_jumpstarting_2016}.
The deepest body of work on this topic was by Sharon Goldberg and Michael
Schapira and their collaborators.  

In 2011, Gill, Schapira, and Goldberg
proposed a strategy that would create market pressure to adopt 
BGP path validation. (They referred to the set of options at the
time as {\em S*BGP}). Their proposal required
(e.g., by regulation) a few Tier 1 ISPs to first deploy S*BGP,
and required those participating in S*BGP to prefer secure routes over
other routes to the same prefix \cite{gill2011-market}.
This scheme also reduced deployment complexity by allowing transit
providers to cryptographically sign routes on behalf of their stub
customers.  Their simulations on realistic AS topologies showed that
under these conditions, the S*BGP ASes would draw traffic away from
other ASes, and most of the rest of ASes would then switch to S*BGP
to get their traffic (revenue) back.  
Followup work two years later \cite{lychev_bgp_2013,CACM-SecureInternet} 
acknowledged that having Tier 1 ISPs lead a market-driven deployment would not work because economic incentive would override any secure route received from a peer when an insecure route via a customer is available.

In 2011, researchers proposed a new Internet architecture,
SCION \cite{scion2011}, that separated ASes into independent
trust domains which provide isolation of routing failures and
human misconfiguration.  Researchers recently used SCION 
to bootstrap a secure routing system \cite{birge-lee-2022}.  
SCION assumes a hierarchical architecture, where one or more 
highly trusted ASes connect the domains to each other.

In 2016, Cohen {\em et al.} \cite{cohen_jumpstarting_2016}
proposed an approach similar to the recently 
proposed ASPA protocol (\S\ref{sec:aspa}).  Their
simulations focused on the length of paths that an attacker must construct
if the AS announcing the prefix has registered what we
today call an ASPA.  The authors discussed deployments
in select geographic regions, perhaps driven by government pressure.
They did not propose a connected region, so partial
deployment of this approach yields only a probabilistic 
assessment of protection, as with ROV (and ASPA).

\subsubsection{Preventing origin hijacks: RPKI and ROV}
\label{sec:background:rpki}

While BGPsec has been undergoing implementation and evaluation for
a decade, operators have focused on the more tractable challenge of
{\em Route Origin Validation} (ROV), which is recognized as the 
best current practice in routing security.  The IETF SIDR WG specified ROV in
2013 as a mechanism to mitigate the risk of {\em origin hijacks} (the
simplest form of hijack)~\cite{scudder_rfc_2013}. ROV uses a Resource
Public Key Infrastructure (RPKI)~\cite{huston_rfc6483_2012}, 
an authoritative database maintained
outside of BGP to store {\em Route Origin Authorization} (ROAs).
ROAs are cryptographic signatures that
authorize designated ASes to originate routes to address blocks (certificates).
Routers using ROV drop BGP
announcements that do not have a matching ROA for the prefix. 
RFC 6811 \cite{scudder_rfc_2013} specifies the ROV protocol 
with important caveats:
its dependence on the integrity of the database used to validate routes,
and its inability to prevent path hijacks. This residual risk includes
the forged-origin path hijack mentioned above, where the malicious AS
impersonates the valid source AS by appending it to a forged BGP
announcement (recently observed in the wild \cite{oliver-imc2022}).  RFC
6811 thus cautioned: ``{\em ..this system should be thought of more as
a protection against misconfiguration than as true `security' in the
strong sense.}''

Use of ROAs presents other operational challenges.
A ROA contains a prefix and a single origin ASN; if an operator wishes to
announce a prefix with different origin ASNs at different
interconnection points, it must issue multiple ROAs.
A ROA may also contain a {\em maxLength} attribute that defines the
maximum prefix length allowed for the prefix; for example, a ROA for
192.0.30.0/23 with a maxLength of 24 enables the AS to originate
192.0.31.0/24. Operators use this feature for traffic
engineering (\S\ref{sec:background:bgp}).
A route is RPKI-valid if any ROA asserts the origin AS in the AS path
is valid.
In 2017, Gilad {\em et al.} showed that use of the maxLength attribute
could enable an attacker to hijack more-specific prefixes
that victim networks then unwittingly communicate with.
Best current practice is to not use the maxLength
attribute~\cite{rpkimaxlen}.

Although RIRs have supported RPKI registration of ROAs since 2013, until
2019 there was little evidence of ISPs using ROAs to validate BGP
announcements.
But by late 2022, many large ISPs, including AT\&T, KPN, Arelion, and Comcast
had started to use ROV to drop invalid
announcements~\cite{kpnfiltering,attfiltering,teliafiltering,comcastfiltering}.
According to NIST's public RPKI monitor based on
RouteViews data~\cite{nist-methods}, as of October 2023,
46\% of IPv4 /24s in unique prefix-origin pairs advertised in BGP were covered
by RPKI and observed as valid, i.e., the origin AS in the BGP announcement
matched the registered ROA.
These statistics vary by region: for 14 October 2023, NIST reported
valid 63\% of observed prefix-origin pairs in the RIPE region, 56\% in
LACNIC, 49\% for APNIC, 32\% for ARIN, and 21\% in the AFRINIC
region~\cite{NIST_rpki_nodate}.
Recent work examining the state of ROV deployment in the Internet
between December 2021 and September 2023 reported that 12.3\% of
tested ASes had behavior suggesting that they or all of their transit
providers had consistently implemented ROV~\cite{rovista}.
They reported that larger ASes (i.e., those networks with technical
capacity) were more likely to have implemented ROV.

\subsubsection{Collective action attempt: MANRS}
\label{subsec:manrs}

In 2014, several network operators established a voluntary initiative
to promote operational practices to ``help reduce the most common
routing threats on the Internet'' -- which they called Mutually Agreed
Norms for Routing Security (MANRS)~\cite{manrs}.
MANRS specified four practices for participating networks, two of
which correspond to the RPKI/ROV steps of registering authoritative
information about one's prefixes, and verifying BGP announcements
against authoritative information.
The exact wording of these two practices are: (1)
{\em Prevent propagation of illegitimate routes from customer networks
or one's own network.}; and (2) {\em Document in a public routing
  registry the prefixes that the AS will originate.}

To conform with the first practice, a MANRS member must verify two aspects of
an announcement from a customer: (1) it must confirm that the customer 
has used an
ASN that it is legitimately allowed to use; and 
(2) for any prefix originated by
that customer, that the ASN is allowed to announce that prefix.
However, to encourage broad uptake, 
MANRS does not specify how a member AS should
verify the assertions of its customers, and in particular does not
require the use of RPKI/ROV (ROAs) in this verification.  The AS can
use ROAs, or can verify against (less authoritative) 
information in the Internet Routing
Registry (IRR), or rely on a private arrangement with its customer.

The MANRS initiative has a key strength: it illustrates that ISPs can
institutionalize their recognition of the need for a collective commitment
to operational practices to reduce threats to the routing system.  However,
as the FCC observed~\cite{fcc-noi-2022}, the MANRS 
program has had limited success.
In May 2023, MANRS had 830 ISP and 22 CDN organizational members that
covered 1011 and 25 ASNs, respectively \cite{manrs-stats}.
This constitutes 1.3\% of the $\approx$75K routed ASes.
Many of the largest ISPs do not participate, and some participating
ISPs are not conforming to the practices.
Du {\em et al.} reported that 5\% of MANRS ISPs did not conform
with the requirement to register their prefixes in
either RPKI or IRR as of May 2022 and 16\% did not conform 
with the filtering requirement ~\cite{du-imc2022}.

The limited success of MANRS (and its underlying practices)
is rooted in misaligned incentives
that manifest in three ways.  First, although if consistently implemented,
the MANRS practices will reduce the incidence of invalid origin hijacks,
{\em there is no direct relationship between the action of any given MANRS
member and the overall security of the Internet, or even the security
of any customer of a MANRS member.}

Second, the current MANRS practices, even the stronger RPKI/ROV
options, only aim to prevent origin hijacks rather than path hijacks. 
Some network operators believe this benefit
does not justify the cost and complexity of RPKI/ROV.

Third, there is insufficient auditing of conformance to lend
confidence in assuming consistent implementation \cite{manrs-conformance}.  
As mentioned, independent auditing has detected
significant non-conformance~\cite{du-imc2022}.  More rigorous
auditing would be expensive and further reduce the incentive
to participate.

\subsubsection{AS Provider Authorization (ASPA)} 
\label{sec:background:aspa}

Recognizing the barriers to BGPsec deployment, and the lack of path
validation capability in ROV, in 2019 several engineers proposed AS Path
Authorization (ASPA) as a mechanism to protect against route leaks
and forged-origin prefix hijacks~\cite{ietf-sidrops-aspa-verification-16}.
As of October 2023, ASPA is still in IETF development. 
ASPA builds on presumed use of RPKI and ROV but enables customer ASes
to go further by registering a list of their transit providers 
in the globally visible RPKI database.  
That database allows any AS to examine a BGP announcement to detect 
and reject many types of invalid path announcements, so long
as the ASes along the path have registered their providers in ASPA.
The authors describe ASPA as preventing route leaks
as well as some forms of path hijacks; it does not prevent an
attacker from spoofing a sequence of ASes in the path if those ASes do
not implement ASPA.  
Our proposed scheme provides a more predictable level of protection 
and improves incentives for deployment. 
We compare our scheme to ASPA, and describe a way in which 
the use of ASPA could expand the
protection provided by both approaches (\S\ref{sec:comparison}).

\subsection{Routing Security in the 2020s}
\label{subsec:2020s}

This decade, routing security caught the attention of regulators. 
Researchers discovered hijacks of unannounced address 
space \cite{ethan-squatting-ccr-2023}, and forged-origin hijacks 
of of RPKI-valid address space \cite{oliver-imc2022}.
After earlier hijacks of AWS address space~\cite{goodin_suspicious_2018}
motivated Amazon to register ROAs for most of its address blocks,
attackers developed more sophisticated path hijacking techniques.
The high-profile hijack of AWS space in August 2022~\cite{ars-hijacks-2022}
motivated by the opportunity to steal cryptocurrency,
succeeded for multiple reasons. Amazon signed multiple ROAs
that allowed different ASNs to originate their prefix; these ROAs had
{\em maxLength} attributes that the attacker exploited to hijack an IPv4/24
that hosted the crypto-currency service; and the attacker registered
that IPv4/24 in an unauthenticated IRR entry to convince upstream
providers to permit the prefix announcement.
However, even if Amazon had announced a competing more specific, the
attacker's path would have been preferred for networks that were customers
of AS1299 who did not have a more-preferred route to Amazon. 

The persistent failure of market-driven solutions to routing security
has recently triggered government interest and 
inquiry into potential interventions.
In 2022, the OECD \cite{oecd-routing}, ICANN \cite{ssac121},  BITAG
\cite{bitag-routing}, and the U.S. FCC \cite{fcc-noi-2022} all
published reports with extensive references related to routing
security challenges, and limitations of proposed solutions.

We expect governments to feel compelled to intervene in the Internet
infrastructure ecosystem to improve routing security, and we seek to
provide an alternative that leaves as much control as possible with the
participating networks. 
Our approach draws inspiration from Lychev {\em et al.}'s 
conclusion a decade ago \cite{lychev_bgp_2013}
regarding market-driven evolution of secure routing:
{\em ``We hope that our work will call attention to the
challenges that arise during partial deployment, and drive the development
of solutions that can help surmount them..  Alternatively, one could
find deployment scenarios that create `islands' of secure ASes
that agree to prioritize security 1st for routes between ASes in the
island; the challenge is to do this without disrupting existing traffic
engineering or business arrangements.''} \cite{lychev_bgp_2013}

We pursue this challenge with an approach that leverages a
coherent topological region to achieve our design goals
(\S\ref{sec:intro}): incentive alignment, competitive
advantage to participating networks; proportional responsibility,
in that larger players can invest to protect their customers,
providing this competitive advantage; and protection against
origin as well as path hijacks without the operational complexity of 
BGPsec. 
We believe our proposed alternative is worth open debate 
before pursuing more blunt regulatory measures.

\section{Threat Model}
\label{sec:threat}

We next describe the capabilities of defenders
(\S\ref{sec:threat:defend}), to contrast defender capabilities with
attacker capabilities (\S\ref{sec:threat:attack}).

\subsection{Defender Capabilities}
\label{sec:threat:defend}

As of October 2023, $\approx$85\% of the $\approx$75K ASes 
are stub ASes that rely on transit
providers and, to a lesser extent, IXPs for Internet routing.
These transit networks engage in contractual
agreements when they interconnect with their neighbors.
These transit network operators regularly interact at peering
forums and other industry
events (e.g., NANOG) and thus have established relationships.
In our threat model, the defenders are these transit providers.
Defenders have the capability to establish parameters with their
customers in terms of what prefix announcements the customer is
expected (and allowed) to make, and thus to automatically accept or
reject routes through configuration capabilities present on routers.
Defenders can access external databases, e.g., IRR, RPKI, 
to support their assessment of
their customer routes.

A defender does not in general have the ability to verify the
announcements of their customers' customers, due to the temporal
dynamism in the interdomain relationships of their customers.
Further, some defenders, and their customers, are limited in how
they use RPKI.
For example, some legacy resource holders are hesitant to obtain ROAs,
as doing so would require they enter a contractual agreement with
an RIR (\S\ref{sec:background:rpki}).\footnote{In particular 
ARIN's agreement embedded a controversial position that in order 
to register a ROA, holders of legacy space, 
(i.e., those allocated before ARIN existed) must contractually agree
that they have no legal property rights to their address space
\cite{yoo_RPKIlegalbarriers_2019}.  In September 2022, ARIN
removed this clause from their Registry Services Agreement
\cite{ARIN-NOI-Jan23}.}
Finally, a defender cannot control the route selection policies of
their peers or customers; these ASes might select hijacked routes
from other neighbors they have.

\subsection{Attacker Capabilities}
\label{sec:threat:attack}

We assume that the attacker controls or has
subverted an AS that connects to the Internet using one or more
transit providers, which provide routing to the rest of the Internet
for that AS and deliver traffic intended for that AS.
The attacker has the ability to corrupt {\em unauthenticated} databases,
such as IRRs, with false claims that they are the legitimate 
holder of a prefix.
Finally, an attacker has the ability to commit to security practices
that they have no intention to follow.

An attacker does not have the ability to completely 
hide their activities; in order for their attack to be
effective, their hijacked route must propagate, and multiple route
collector projects today publish a comprehensive set of AS paths. 
Nor does an attacker have the ability to issue ROAs for address space
that they do not control, unless they compromise the RIR (an insider)
or the prefix holder's RIR account.

\section{A Routing Zone of Trust}
\label{sec:design}

\begin{figure}[t]
\centering
\includegraphics[width=0.8\linewidth]{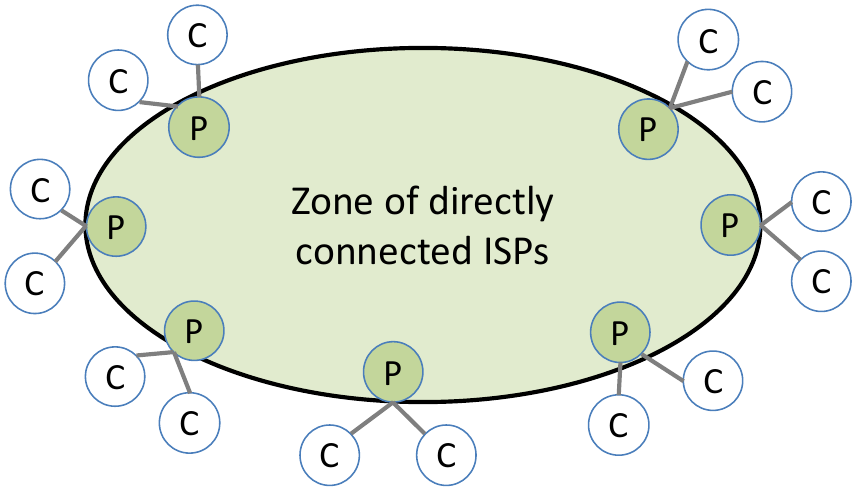}
\caption{One requirement for our conceptual zone of trust is a coherent topological region with providers in the zone providing transit to customers attached to those providers.}
\label{fig:simple}
\end{figure}

We first introduce the concept of a zone of trust in a routing context
before specifying one in more detail in \S\ref{sec:design:vipzone}.
Figure~\ref{fig:simple} depicts a zone with member 
providers (in green) at the edge of the zone providing transit
service to directly attached customers (white). The providers
connect within the zone,  and must know when they are exchanging
traffic with another member of the zone, and when they are
communicating with an AS outside the zone.

A zone could protect against origin 
hijacks as follows. If all providers P in the zone commit to implement ROV and drop invalid announcements from customers outside the zone, then no invalid announcements will circulate inside the zone, which means that customers C will never receive a 
BGP announcement from the zone where the origin is invalid based on a ROA. 
These practices turn this zone into a {\em zone of trust}. 

This example illustrates three properties of a zone of trust:
\begin{itemize}
\item \emph{Collective action} by ASes creates the zone and its trust attributes. 
\item The trust attributes of the zone depend on the topology of members within the zone
\item Customers of the zone obtain protection by using a provider in the zone. They need take no other action. 
\end{itemize}

We call this region a \textit{zone of trust} because
the protection arises at the perimeter of the zone.  This protection
requires that ASes in the zone be able to trust that the routers at the
perimeter function correctly, which requires some degree of transparency
and accountability.    We introduce this design assumption in exchange
for one that routing security protocols have always included: 
global deployment of a protocol.  If ASes themselves are threat actors,
we are skeptical of an aspiration to make BGP {\em globally}
secure.  Creating a zone of trust through perimeter protection
(a trust-but-verify regime) offers a more pragmatic approach
for today's routing system.

The idea of a coherent perimeter around a zone is missing from today's
interdomain routing system. ROV deployment discussions today consider   
each AS in isolation, leaving security a statistical measure.
We can count the number of ASes that register their ROAs,
or the number of ASes that implement ROV, but the consequence for a given
AS is a function of what other ASes choose to do.  It is thus not clear
what specific action an AS should take to reduce its own risk profile.
Today, invalid announcements may propagate across the Internet,
and may or may not reach any given AS.  In contrast, a connected zone of trust
allows clear articulation of the benefit to a given AS to joining
the zone: ASes in the zone will receive no announcements from the zone
with an invalid origin based on a registered ROA.

The incentive alignment extends beyond the zone: customers
concerned about hijacks can seek out providers that are in
the zone, which in turn creates an incentive for providers to
commit to the required practices that define the zone and join
it. Today, there is little direct benefit to an AS that chooses
to implement ROV.  Many of the larger ASes do so, as part of a
collective action to improve security, but recognizing that
these actions can create a coherent zone with direct benefit
to their customers will increase their incentive.

Note that the zone does not provide absolute protection from
origin hijacks.  If a customer C has its own customers, peers,
or other providers not in the zone, it could still receive a
hijack from those nearby ASes. We call this set of ASes the
\emph{local region} of the customer C, and we characterize
this residual risk in Section~\ref{sec:considerations}.
Importantly, the residual risk depends on the size and character of the
local region of each AS, which they can know and control
according to their own risk profile. 

Figure \ref{fig:simple} allows us to consider how the trust zone
gives each AS control over the two types of hijack harm:
having one's addresses hijacked or having one's traffic hijacked.  
An AS (e.g., B) can protect against hijacking of its own addresses
(which we call {\em owner harm}) in the zone by directly
connecting to the zone, and registering its addresses in
the RPKI (\S\ref{sec:considerations}).  
Other ASes that attach to the zone are thus protected from
hijacking of their traffic to B's addresses (which we call
{\em misdirection harm}.  An AS that does not consider owner
harm a significant risk need not register its addresses in a
database (although we encourage universal use of the RPKI).
The AS may care more about misdirection harm and might thus
minimize its local region and get as many route announcements
as possible from providers in the zone.  Different ASes may
have different risk assessments, and unlike today's routing
ecosystem, this trust zone is structured to allow an AS to pick its own
options based on its own assessment of route hijack risk.

\paragraph{Does a coherent zone exist?}
\label{subsec:manrs-zone}

Could such a coherent topological region exist?
In fact, it already does, in the context of the MANRS initiative.
Many of the MANRS members make up a connected region today. 
In May 2023, MANRS had 830 ISP members of MANRS, with 1011
ASNs~\cite{manrs-stats}.  To derive the connected region, we
perform a topology exploration using the CAIDA ASrank
data~\cite{asrank-data} for May 2023.
We start with members with no providers (Tier 1 providers), and
recursively add directly-connected customers that are also MANRS members.
The resulting region has 499 members with 613 ASNs. 
Currently 25,916 customers directly connect to this region.
If MANRS could extend their operational practices to make this
region a zone of trust, approximately one-third of the ASes active
on the Internet today would receive that protection.

\begin{figure}
\centering
\includegraphics{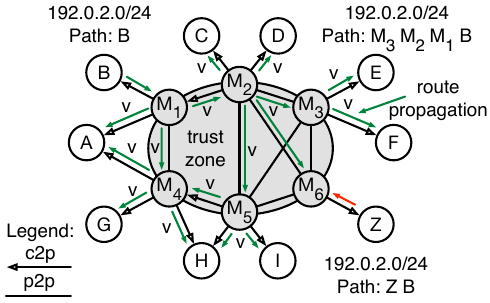}
\caption{
A Routing Zone of Trust can defend members and their customers from
path hijacks in the zone if members (M) mark routes from their
customers as~\ver~(v) as they enter the zone, and other zone members
select~\ver~routes over unverified routes.
Above, M$_{1}$ expects its direct customer B to announce
192.0.2.0/24, so M$_{1}$ marks that route as~\ver, and propagates it
to other members.
Lines with hollow arrows show c2p links, lines without arrows show p2p
links, and lines with solid arrows show route propagation.
The hijacked route via Z does not propagate in the zone, because Z is
not a member, and the zone has an alternative~\ver~route.
}
\label{fig:pzone}
\end{figure}

\subsection{Verified IP Zone (VIPzone)}
\label{sec:design:vipzone}

We now describe how a set of proposed operational practices
in a coherent zone of trust, which we call {\em VIPzone}
(for Verified IP zone), will limit
path hijacks.  For an AS to be in the VIPzone, it must commit to these
practices, and must be part of a connected zone.  To be part
of the connected zone, it must either be a Tier-1 provider,
or have a member of the VIPzone as a transit provider.

Figure~\ref{fig:pzone} illustrates the basic VIPzone operation.
We describe the VIPzone practices below, and provide a
finer-grained specification of these practices in \S\ref{sec:rules}.

Our {\bf VIPzone} builds directly on MANRS requirements 
(\S\ref{subsec:manrs}) \cite{manrs}. 
Each MANRS member is required to verify all
announcements originated by its directly connected customers.
The member must perform two checks: (1) 
that the customer has used an ASN that it is
legitimately allowed to use, and 
(2) for any prefix originated by
that customer, the ASN is allowed to announce that prefix.
The member must rely on direct knowledge of its customer (a
``know your customer'' or KYC requirement) to verify that the
AS used is legitimate.\footnote{BGPsec (\S\ref{subsec:bgpsec}), which 
has not achieved any
material deployment, is the only technical mechanism that can
confirm that a customer controls the AS that it is using for
its announcements.}  
A (MANRS or VIPzone) member can use RPKI validation, an
authenticated IRR database, or a manually-configured prefix
list (ACL) to verify the non-member's announced prefix is
correct.  
In our VIPzone scheme, the zone member then either drops
such announcements  or marks them \ver. We propose the use of
a community value~\cite{rfc1997} to carry the \ver~marking,
similar to a recent IETF proposal to use a community value 
to annotate path properties in order to allow detection of route leaks
\cite{ietf-grow-route-leak-detection-mitigation-09}.
For announcements that come from customers of the customer,
the VIPzone member forwards them without marking them \ver. 
Neither MANRS nor VIPzone requires that an AS check the validity of 
the path in an announcement with more than one AS in the path.

However, for announcements that VIPzone members have verified,
they must propagate the \ver~marking as they forward announcements 
within the zone. A member must remove this
marking if it appears in any announcement entering from outside
the zone.  This allows VIPzone members to establish the
authenticity of \ver~announcement,
regardless of their distance from the origin.
Finally, inside the zone, any AS receiving multiple announced routes for
the same prefix must prefer one marked \ver.
By this rule, no member will prefer a path hijack route over a
legitimate route from customers directly attached to the zone,
since legitimate routes will be marked \ver. 

Customers directly connected to the zone minimize owner harm, 
both for origin and path hijacks. Zone members verify prefixes
received from attached (non-zone) customers and then forward them 
into the zone marked \ver. If a malicious
AS directly connected to the zone tries to launch an invalid
origin hijack, zone members will discard it based on the 
KYC practices. If the AS launches a path hijack (which must by
definition have more than one AS in the path), the member AS
may forward it unverified into the zone (a ``not sure'' situation), 
but it will have no impact so long
as a corresponding \ver~announcement is active.

We emphasize the essential role of the~\ver~tag.
When a MANRS member cannot verify whether
the path announcement is valid (e.g., multiple ASes in path) 
the member can forward this announcement onward.  
Forwarding potentially invalid announcements without {\em any} signal of
risk prevents the current MANRS framework from manifesting a zone of trust.
A key requirement of the VIPzone approach is that members propagate
two sorts of announcements in the zone: \ver~and ``not sure''.
The feature allows for more flexible and incremental
deployment of the protections. In our VIPzone proposal, each AS
drops invalid announcements, marks announcements as \ver~if it knows they
are correct, and forwards announcements without the \ver~marking if the AS is ``not
sure.'' The rule that makes the zone trustworthy in this case is
that if there is a \ver~announcement for a particular prefix, and
one that is not \ver~(e.g., ``not sure'') for the \emph{same prefix},
the zone members must prefer the \ver~announcement.
This rule constrains the routing policies of zone members, the 
implications of which we discuss in \S\ref{sec:considerations} and 
elaborate on more complex scenarios in \S\ref{app:hijacks}.

\paragraph{Protection against route leaks}
A route leak is an event in which an AS inappropriately (i.e.,
violating routing policy) forwards a route it legitimately received.
The consequence is often that large flows of traffic reach this
AS, which is not provisioned to carry them.  A classic
route leak occurs when a multi-homed AS that takes the routes
it receives from one of its transit providers and inadvertently
propagates these routes to its other transit provider.

In addition to preventing path hijacks of ASes directly attached
to the zone, the VIPzone prevents leaks of announcements of prefixes
by ASes \textit{not} in the VIPzone. If the leak occurs within the 
zone, the announcement would be \ver~and thus propagated within the zone.
This potential harm from accidental misconfiguration suggests
an important insight: most ASes should not be in the VIPzone,
but should get the protections by being a \textit{customer}
of a VIPzone member.  We consider it preferable that only
operators with sufficient technical abilities and resources
join the VIPzone.  We elaborate on this idea in \S\ref{sec:leaks}.

\paragraph{Protection against sub-prefix hijacks}
One hijack that can penetrate the zone is based on a sub-prefix (an
address block that is a subset of a \ver~prefix).  Normal routing rules
require that an AS, when selecting among routes for an arriving packet,
must prefer the announcement with the longer prefix (i. e., smaller
address block).  Note that requiring that a \ver~announcement for a given
prefix take precedence over an un\ver~announcement for a longer prefix
risks breaking traffic management practices that disaggregate
prefixes. Such a requirement could introduce loops.  An AS
concerned about owner harm resulting from a sub-prefix attack
protects itself by registering ROAs for the prefix.

\begin{comment}
\ddc{As per kc, consider moving or deleting. Do we need this level of detail?}
However, the
degree of risk mitigation depends on how it configures ROAs. The
ROA option \emph{max length} allows a prefix owner to register a
ROA that allows a range of valid prefix lengths in announcements.
As we discussed in \S\ref{sec:background:rpki}, an AS using this
option is trading off increased owner risk of a sub-prefix hijack
in the zone in exchange for flexibility in how it announces its
addresses. The benefit of this flexibility depends on how easy it
is for an AS to register new ROAs, and how rapidly they propagate
through the Internet. But the choice is up to the AS.
\end{comment}

\subsection{Evaluating VIPzone protections} 
\label{subsec:evaluation}

\begin{figure}
\centering
\includegraphics[width=0.8\linewidth]{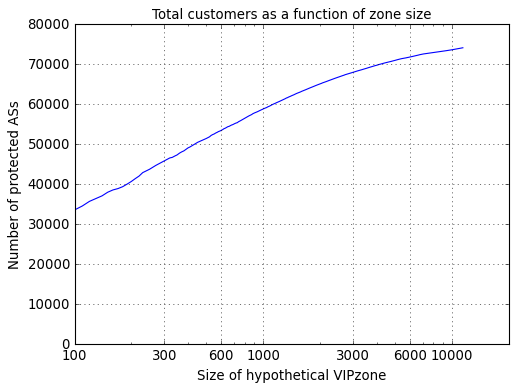}
\caption{Protected ASes (in the zone or connected directly to it) as a function of zone size (ASRank data, May 2023)} 
\label{fig:psizes}
\end{figure}

We explore how many ASes would receive protection from hypothetical
zones based on today's Internet topology.  
Using CAIDA's AS Rank data from May 2023, 
we initialize a zone with the 100 ASes with the largest customer cones.
We then add new members, again ordered by the size
of their customer cone.  Figure \ref{fig:psizes} shows the
number of protected ASes expanding rapidly with zone size, up
to 11,458 ASes (in this data set), at which point every AS with
any customers is in the zone.
The only ASes not in the zone are single AS stubs. 

However, note that such a large zone is unrealistic. Most ASes
in that zone are small providers with few customers, likely
without sufficient operational sophistication or resources to
join the zone. If we pick an arbitrary cutoff of 600 members
(about the size of the current MANRS zone), would protect
a little over two thirds of the ASes in the Internet (in this
hypothetical analysis, 53,112). 
This number is higher than the 25,916 customers of the current
MANRS region (\S\ref{subsec:manrs-zone}), because this VIPzone
is formed by including all of the largest ASes (even non-MANRS
members) as measured by
their customer cone.

\section{Evaluating Residual Risk (Local Regions)}
\label{sec:considerations} 

We review the residual risks that an AS faces even if it is
directly connected to the VIPzone, and what that AS can do to
further reduce these risks.
We have already described how an AS mitigates the risk of
\emph{owner harm} by connecting to the zone.
The residual risk of {\em misdirection harm} if they connect to
the zone is a function of the size of the {\em local region} and the 
probability that a malicious AS operates in that region.  

\begin{figure} \centering
\includegraphics[width=0.8\linewidth]{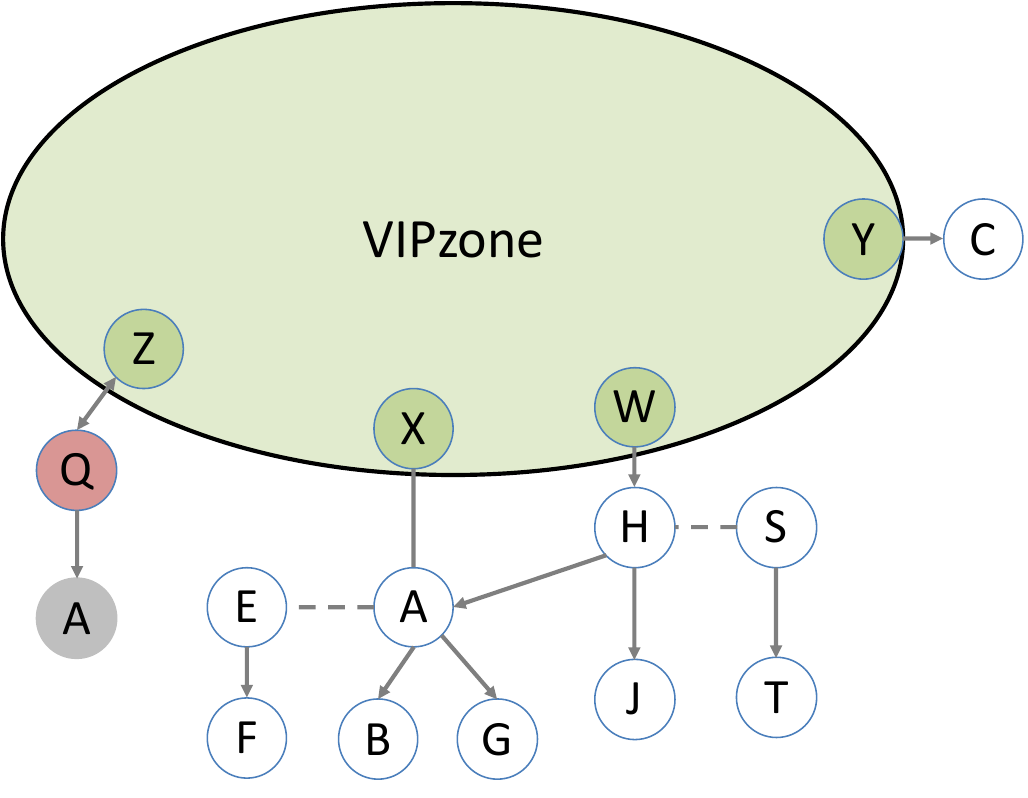} 
\caption{Various customers of a VIPzone, including A with 
	a local region, C with no local region, and 
	a malicious AS Q pretending that A is a customer.} 
\label{fig:local} 
\end{figure}

A local region of a VIPzone customer arises due to its interconnection
arrangements outside the zone, from which it receives BGP announcements.
These include the ASes in the customer cone of that AS, the peers of
that AS and their customer cones, and any providers (and their
neighbors, recursively) of that AS that are not in the zone.
In Figure~\ref{fig:local},
A has provider X in the zone. Its local region includes customers B
and G, peer E and E's customer F, provider H (which is not in the zone), and its customer J, and peer S of 
provider H and its customer T. If provider H itself had a provider that was not in the zone, that provider, its customers, and any peers and customers of those peers would also be in the local region of A. Any of these could launch a hijack
that triggers a \emph{misdirection harm} to A.

We make three observations about local regions. 
First, the risk of hijack by one's own customer 
(A's customer B in this example) is a function 
of the risk of malicious behavior in the local region. 
But A (or any AS outside the zone) can mitigate this risk 
by implementing a robust KYC practice, which can generally 
detect forged-origin attacks by customers.

Second, misdirection from a hijack in the region is restricted
to the region.  
In Figure~\ref{fig:local}, if malicious AS Q launches 
a path hijack asserting that it has A as a customer, that
announcement may penetrate the zone but without a \ver~mark,
so zone members will prefer the \ver~announcement from X. 

Third, for many attached customers the local region is small.  To
examine the size distribution of local regions, we return to our
hypothetical VIPzone (i.e., seeded with 100 ASes with the largest
customer cone) and compute the size of the local regions for all
attached customers.  We add to the zone 100 ASes at a time, and at
each step compute the size of the local region for the attached customers.

\begin{figure}[htb]
\centering
\setkeys{Gin}{width=\linewidth}
\begin{subfigure}{0.4\textwidth}
\includegraphics{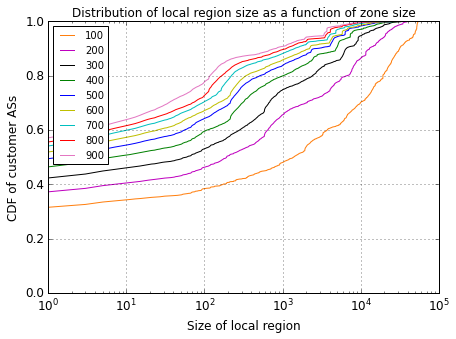}
\caption{Plot based on data from ASrank}
\label{fig:lr-nopeer}
\end{subfigure}% trailing space between `subfigure` environments  had to be removed
\hfil
\begin{subfigure}{0.4\textwidth}
\includegraphics{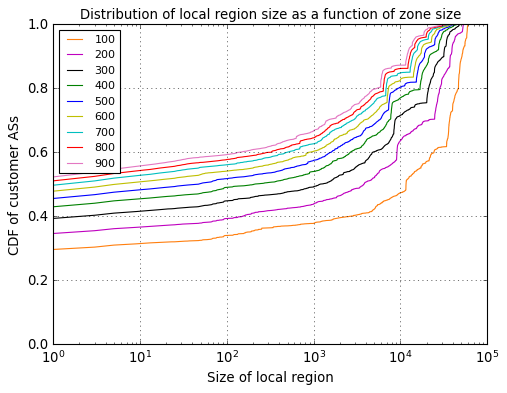}%{diagram.pdf}
\caption{Plot based on data from ASrank plus an assumed peering relationship between any two ASes at an IX}
\label{fig:lr-allpeer}
\end{subfigure}
\caption{Sizes of local regions for customers of a hypothetical
VIPzone, for various zone sizes. Between 30\% and 40\% of the customer ASes have a region size near 1.}
\label{fig:lrsizes}
\end{figure}

Figure \ref{fig:lrsizes} plots the resulting distribution. 
To compensate for the limited observability of peering relationships, 
we use two methods to compute the size of the local region. 
Figure~\ref{fig:lr-nopeer} plots the local region size using the customer-provider  and peering relationships from CAIDA's ASRank data \cite{asrank}. Figure~\ref{fig:lr-allpeer} relies on the method in~\cite{lychev_bgp_2013}, which 
augments observable peering relationships by assuming that any two ASes that attach to the same IX have a peering relationship. We use data from from 
PeeringDB and PCH \cite{caida-ixpdata} to augment the set of
peering relationships inferred by AS Rank.  

Figure~\ref{fig:lr-nopeer} underestimates the sizes
of local regions, since CAIDA's ASRank
data is derived from BGP announcements collected by RouteViews
and RIPE RIS \cite{routeviews,RIPE-RIS}, and those vantage
points do not have sufficient density to capture all peering
relationshops. Figure~\ref{fig:lr-allpeer} probably overestimates
the sizes of the local regions, since many ASes that connect
to IXs have selective peering policies. So the actual distribution
probably lies between these two set of curves.

Note that the distribution of local region sizes is bimodal. Depending
on the zone size, between 30\% and 60\% of the customer ASes have a
very small local region--close to 1 AS. These are stub ASes
that obtain access to the Internet using a transit provider, and
do not peer to obtain connectivity. The right side of the plots shows
large local regions, which represent ASes that peer widely
to reduce their dependency on transit providers, or
else use multiple providers, one of which is not in the zone, and which
itself uses massive peering. A realistic consequence of 
extensive peering with ASes that 
do not take known steps to verify their announcements is an increased
risk of hijack. That expanded attack surface in the ecosystem
is a motivation for the approach we propose.

A further uncertainty in these plots derives from the common use of
prefix filters on peering links, precisely to protect themselves from
harm due to erroneous or malicious BGP announcements. That practice
would reduce the effective size of the local region from which hijacks
can come. Many ASes consider such filtering good routing hygiene today.
We know no way to measure the extent to which operators have
deployed such filters.

\paragraph{Protection for ASes not attached to the zone} 

In Figure~\ref{fig:local}, AS B shares the local region of A, but is not
directly connected to the zone. What protection does B receive
from hijacks? With respect to owner risk, B can prevent simple hijacks
based on an invalid origin by registering ROAs, but it gets no protection
from path hijacks. With respect to misdirection risk, it is in 
the same situation as A: no hijacks will come into B's region from the
zone, but a hijack in B's local region can still cause misdirection harm.
Many smaller ASes offer low-value, limited-interest services,
and their owner risk of a hijack is minimal. If the AS does
consider the owner risk to be substantial, they can and should
obtain transit from a member of the zone.

\begin{comment}
\paragraph{Residual Risk} As protection against traditional hijacks improves, attackers devise new ways to disrupt routing. One is a social engineering attack in which an attacker contacts a provider of a target AS, and (pretending to be an agent of the target AS) requests that the provider provision a new link to serve that target AS. If the provider does not recognize that the request is not legitimate, the attacker now has a BGP connection to the provider that the provider thinks is associated with the target AS. At this point, the attacker can announce routes (e.g., hijack them) associated with the target AS, and the provider will accept these announcements.  \mjl{how does this impact the victim?}

Transit providers will have to harden their implementation of the MANRS Know Your Customer requirement to detect these sorts of attacks. This requirement applies equally to the existing MANRS, VIPzone, and ASPA (\S~\ref{sec:aspa}).
\end{comment}

\paragraph{Why we cannot assess realistic risk using current topology data.}
This analysis provides a hypothetical indication of the 
level of protection and residual risk that a VIPzone 
would yield under current interconnection patterns. 
But it is a problematic approach to assessing residual risk,
since the architecture of the VIPzone will affect peering
incentives, by design.   That is, the goal of VIPzone proposal 
is to devise
a set of practices that allow an AS concerned about security
risk (in particular the risk of hijack) to take action that 
minimizes this risk. In other words, they will shift 
interconnection patterns to exploit the benefit of the zone.

ASes not in the zone who want protection
from hijacks have several options: assessing their
peering partners as to the risk of malicious or erroneous
behavior, installing appropriate prefix filters, disconnecting
from low-traffic peers.  Some peers may take steps to verify
their own customers, and the practical risk of using routes
from such a peer would be minimal.\footnote{Internet2 exemplifies
such a region; they track the full customer cone of their
members, and use prefix filters to prevent incorrect announcements.
Using routes from a region of this sort is practically risk-free.}
But we emphasize that the power of a trust zone approach is that
each AS gets to make its own risk assessment, and
act accordingly.  ASes with small or no local region would not
have to take these steps.  Larger ASes are more likely to have
the operational capacity to protect their local region,
e.g., implement prefix filters. If a
VIPzone existed, we would expect ASes to take actions
to reduce the residual risks from their local regions.

\section{Auditing Requirements} \label{sec:auditing}

Our proposal for a VIPzone does not use real-time detection
of suspicious announcements.
Real time prevention requires adding code to the BGP processing
path in routers or route computation servers. This approach
would potentially lead to a more brittle scheme.
Instead the VIPzone uses a trust-but-verify approach:
checking conformance of members with its requirements, 
detection and documenting of failures, and
suspension or ejection of non-compliant members.  
This requirement means that members must have the will, and the 
institution, to undertake conformance auditing.
Independent third parties can check conformance off-path,
by looking at public BGP announcements. In support of this auditing,
every VIPzone member would be required to provide a BGP view to a route
collector.  The audit process does not use the member's view
to audit their behavior (the member could lie) but rather 
uses the views provided by the member's neighbors that are also
members and thus provide views of their own.  Using the neighbor views
allows confirmation that the member correctly propagated verified routes
with the \ver~tag, and did not use the \ver~tag on routes that other
members had not tagged as~\ver.

This approach is similar in spirit to how the CA/Browser forum verifies
the correct behavior of certificate authorities.
Its goal is not to detect and block every issuance of a false
certificate in real time, but rather to identify CAs that are shown to
be untrustworthy and remove them from the list of trusted root CAs
included with distributions of browsers.  The idea is to enforce proper
behavior by making the consequence of misbehavior a substantial penalty.
In that context, the CA/Browser community has shown a willingness to
take action against providers that do not conform.  For the VIPzone to
provide protection in practice, the routing community must have the same
will. We argue that an industry-led body, analogous to the CA/Browser
forum, should decide on necessary actions if a VIPzone member does not
conform to the required practices.  But note that the
penalty in this case is not being disconnected from the Internet, but
just losing the right to initiate \ver~announcements.

Independent of the exact specification of the practices that define a zone, it must be possible to tell by inspection if an announcement is not conformant.
The three tests for VIPzone member conformance are:

\begin{itemize} 

\item Rule 1: If an announcement (observed anywhere in the
VIPzone) has more than one AS number in the path before it enters the VIPzone,
and is marked VERIFIED, the member that introduced the announcement into the
core is non-conformant. Our trust model assumes that verification and checking
of announcements occurs at specific locations: the ASes at the edge of the
VIPzone that have customers not in the VIPzone.  This requirement makes it
possible to identify members that do not implement the required practices.

\item Rule 2: If an announcement has an invalid origin, as determined by a ROA,
independent of path length, the VIPzone member that introduced the announcement
is non-conformant.
  
\item Rule 3: ASes in the VIPzone must forward the \ver~ community value from
other VIPzone members.

\end{itemize}

Another advantage of off-line conformance checking is that it could
allow an AS to register its intent to announce a route that is non-conformant
(e.g., to deal with a specific customer requirement), and accept
responsibility for ensuring that it is benign. Allowing benign exceptions,
including marking them \ver, enables more nuanced balance of security 
and availability priorities. 

\begin{comment}
A challenge with conformance checking is to check proper implementation of the
required practices without waiting to receive an invalid route. To see if a
VIPzone member is removing the \ver~value from an incoming announcement from a
non-member, it would be necessary to send such an announcement to the member
from a connected AS that is not in the VIPzone, and observe what happens to the
community value. \kc{discuss: Because this test involves the control plane,
rather than the data plane, a third-party cannot inject such a routing
announcement  as part of the test.}
\end{comment}

\paragraph{Cost of checking for conformance}  The conformance checking
requirements imply non-trivial costs.   The data collection and curation infrastructure would
require staffing to maintain, whether operated by an independent
private-sector group such as RouteViews, or some more formally chartered
institution or agency. Then one or more technically capable
organizations must perform the auditing and 
provide the information necessary to judge
untrustworthy behavior.   

\section{Comparison to Other Proposed Solutions}
\label{sec:comparison}

We compare our proposal to two leading alternative proposals
to advance the collective state of routing security, in particular
to prevent path hijacks: BGPsec and ASPA.  
But we preface this comparison with a comment on the 
tension between our VIPzone approach and the philosophy
of {\em zero trust architectures}.  Zero trust is usually proposed in a
context where each machine or subsystem performs its own verification
to protect itself, and the incentives are directly aligned
\cite{nist-zerotrust}.  The collective action aspect of routing security,
where it is not feasible to verify implementation by other parties,
is at odds with this assumption.  

The VIPzone approach better aligns incentives, allocates
responsibility to specific points in the zone (the perimeter),
and ascertains whether zone members are implementing the
required operational practices.

\subsection{AS Provider Authorization (ASPA)}
\label{sec:aspa}

ASPA~\cite{ietf-sidrops-aspa-verification-16} is a mechanism that lets
a customer AS register a list of providers that the customer uses. This
registration (an Autonomous System Provider Authorization or ASPA) is
recorded in the same system that is used to store ROAs--the RPKI
administered by the five RIRs. The ASPA data is globally visible, so any
AS receiving a BGP announcement can look at the sequence of ASes in the
path, and check to see if there is an ASPA that covers any adjacent pair
of ASes in the path. If there is, and the announcement is inconsistent
with the ASPA, the AS receiving the announcement can drop
it~\cite{ietf-sidrops-aspa-verification-16}.
ASPA can be used to limit both route leaks and, to some degree,
against path hijacks, assuming the appropriate ASes deploy ASPA
in the correct places.  The ASPA specification describes 
several deployment scenarios.

ASPA's design differs in several ways from our proposal.

\begin{itemize}

\item The VIPzone design tries to minimize the effort required of small ASes to get protection. It requires only that the small AS connect to a transit provider that is in the zone and (ideally) register its ROAs. ASPA  requires that the small AS register an ASPA describing its providers. While the mechanics of registration need not be complex, this registration becomes one more data record that the operators of the AS must keep track of, and remember to change if they change providers.

\item The VIPzone design does not require new mechanisms
in the routers (or route computation servers). 
The actions required of a VIPzone member (\S\ref{sec:design:vipzone})
include new operational practices and use of a new community value.
ASPA checking requires a new processing check, which includes downloading
the relevant ASPA data and inspecting the announcement for validity.
This dependency also implies the need for the RPKI to 
store and manage new (ASPA) records.

\item The VIPzone design assigns clear responsibilities:
an AS at the edge of the zone has specific requirements to 
check announcements received from its customers, including
a KYC check.  This perimeter 
allows clear description of protection and residual harm. 
The current ASPA draft~\cite{ietf-sidrops-aspa-verification-16} 
describes use cases without assigning responsibilities to
specific ASes. Thus it is not clear which ASes should do ASPA
checking, which ASes would have the motivation to register ASPA
records, and (thus) what protection ASPA will achieve.
For example, if an AS has listed a provider in an ASPA record,
and that provider has such poor business/operational
practices that it cannot identify an imposter posing as their legitimate
customer, an ASPA alone cannot prevent the resulting harm.
Assignment of responsibility, as in VIPzone, allows the possibility of
conformance checking.

\end{itemize}

\begin{figure}
\centering
\includegraphics[width=0.8\linewidth]{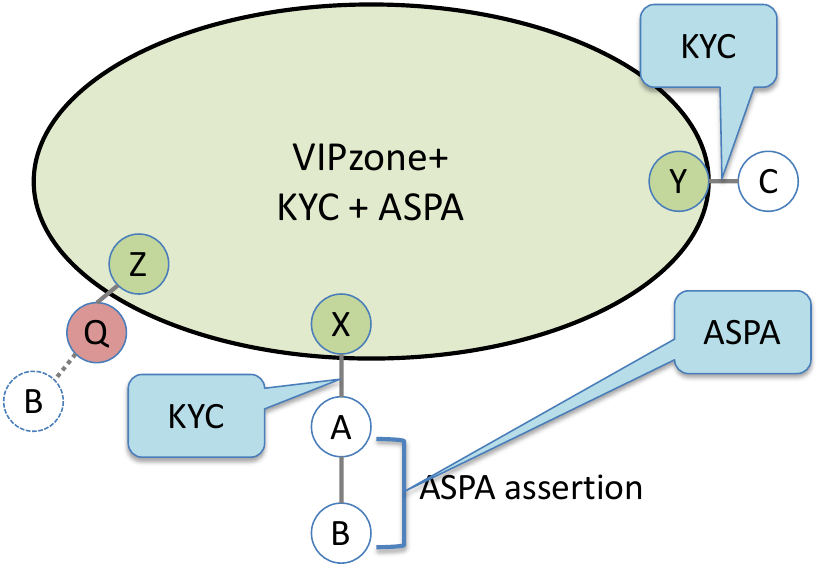}
\caption{{\bf ASPA-extended VIPzone.}
Zone member X can
use ASPA to verify an announcement with two ASes in
the path outside the zone. If B registers an ASPA recording
A as a provider, and X has done KYC on A, then X can mark
the route as VERIFIED.}
\label{fig:aspa}
\end{figure}

ISPs could use ASPA to further the range of VIPzone protection
to customers of zone customers 
This extension would allow an AS at the
edge of a zone to mark as \ver~ announcements with two or fewer ASes in
the path, as opposed only one (\S\ref{sec:design:vipzone}).
In Figure \ref{fig:aspa}, Y uses VIPzone practices to verify
announcements originated by C, as does X to verify A. But
X has the option of using an ASPA registered by B to confirm
that A is a valid provider of B.
X can tag as \ver~the announcement that includes both A and B in the path only if there is an ASPA registered by B. Otherwise, X must not mark the announcement, but can forward it into the zone unmarked.

The other case in Figure~\ref{fig:aspa} is that Q is malicious, and
wants to hijack a prefix belonging to B.
If B has registered a ROA for the prefix, then Q
cannot validly announce B's prefix. It would have to pretend to be B.
If B has registered an ASPA saying that its provider is A, then this
ASPA would allow Z to conclude that the announcement is invalid. The
attacker Q could add  AS A to the path to make a valid path, but then
the announcement would have three ASes in the path outside the zone
(A, B, Q), and
(in the VIPzone we propose) Z must not mark a path \ver~ if it has
more than two ASes in the announcement.

\subsection{BGPSEC} 
\label{subsec:bgpsec}
Like ASPA, BGPsec is attempting to achieve a zero-trust approach.  To
the extent that every router that forwards the announcement adds its own
cryptographic signature, any router along the path can verify that the
series of signatures to that point are valid.  This function also means
that BGPsec, if pervasively and correctly deployed, provides the technical
means to address the KYC requirement that VIPZone and ASPA cannot do
in-protocool.  That is, BGPsec prevents
the social engineering impersonation attack, 
since the imposter will not have the necessary keys to sign
their announcements.  However, the requirement for comprehensive 
deployment dramatically reduces the incentive for ISPs to undertake the cost
and complexity of BGPsec deployment, and a lengthy trajectory of
partial deployment implies inconsistent and unpredictable implementation
of the required checking.  We expect that governments will not
have the patience to wait for deployment of a global solution
to route hijacks.

\section{Conclusion}
\label{sec:conclusion}

There is currently no consensus as to the next step to secure BGP beyond
the simplest type of hijacks.  As of 2023, BGPsec has no production
deployment, and arouses significant controversy over the operational
feasibility of its key management aspects.  For all proposed solutions
to prevent path hijacks, incentives are misaligned.  We have proposed a
path forward that creates incentives for ASes (both customer and
provider) to participate, protects ASes against {\em path hijacks}
and {\em origin hijacks} with no effort or investment needed by small
ASes, and avoids the need for new mechanism in routers.

One insight that shapes our proposal
is that if there is a coherent topological region of the Internet, 
and with practices limiting malicious BGP routes entering that region,
then the operational practices can provide much stronger
protection against abuse for those who join, and thus incentive to
participate.  The result is a virtuous circle, where customers benefit
from choosing ISPs committed to the practices, and ISPs (thus) benefit
from committing to the practices.
A coherent core of ISPs has already emerged
organically in the ecosystem, which can be leveraged to create
a {\em zone of trust}, a region that protects not only all
networks in the region, but all directly attached customers.

A few concerns with VIPzone bear further consideration.  First, will it
concentrate power in a few trusted networks, those with the authority
to verify routes?  We believe the VIPzone requirement for transparency,
accountability, and independent auditing, provides a counterpoint to
potential abuses of power.

Second, will trust zones fragment the Internet?  Some Internet 
fragmentation has already occurred, and trust zones provide a way to
bridge some of these fragments using a trust-but-verify framework, like
treaties in other global domains.   We acknowledge that multiple trust
zones may emerge, including on national boundaries.  
But note that for a VIPzone to be effective, both (1) ASes that produce important services, and (2) ASes that consume those important services, must be attached to that zone.  As an extreme example, if each country
wants its own trust zone, networks with global customer bases would have
to replicate their point of attachments in all trust zones where they
serve customers.  We imagine trust zones to evolve instead more like global
trading zones.

Third, achieving the VIPzone protection requires auditing and 
enforcing conformance with the practices. The institutional framework
required for such checking already exists in multiple places,
e.g., RIPE and RouteViews.  But it is still more expensive (and therefore
less incentive-compatible) than doing nothing in the current unregulated
environment. 

Fourth, ISPs have to trade off some autonomy in exchange for routing
security. ISPs are required to prefer VERIFIED routes over customer
routes, and ISPs would hand some control over to a non-ISP third party
(the auditor) similar to the CA/Browser Forum today.  But unlike other
proposed approaches to routing security, transit ISPs can claim to 
offer their customers a securely-routed service by participating. 

Our proposal responds to a long-standing need for some medium-term path
forward on protection against path hijacks. We believe it is a direction
worth debate and analysis in the context of possible regulatory measures.  
We recognize that ISPs, like most private sector actors,
prefer lack of regulation and work to avoid it as long as possible.
But the EU has made it clear they will regulate to safeguard their
citizens despite private sector objections \cite{gdpr,eudsa,eucra}.
We offer this path forward as an approach where the private sector
could drive a self-regulatory framework that achieves the accountability
regulators are now seeking in digital domains.

\bibliographystyle{plain}%abbrv}
\bibliography{bib}
\begin{appendix}
       { \centering\section*{APPENDIX}}
\section{Full specification of required actions for members of the VIPzone}
\label{sec:rules}

We summarize the required operational practices of VIPZone
members in \S\ref{sec:design:vipzone}; here we repeat the
summary, provide additional details, and diagram specific scenarios. 

First, VIPzone members that can participate in these enhanced
practices must be part of a connected region.

Second, if a VIPzone member receives a BGP announcement from a
neighbor that is not in the zone, and the announcement is for a prefix
that the neighbor \textit{originates} and the member can verify as
legitimate, then the member will tag the route with a new BGP
community value \cite{rfc1997}, which we call {\em \ver}. (Some other BGP mechanism with 
equivalent properties could also be used.)

Third, VIPzone members must propagate this community value as they
forward announcements to other ASes.
This allows neighbors to establish the authenticity of the route,
regardless of the distance they are from the origin.

Fourth, inside the zone, any AS receiving multiple announcements for
the same prefix must prefer one marked \ver.
By this rule, no member will prefer a path hijack announcement over a
legitimate announcement from customers directly attached to the zone,
since those will be marked \ver.

The operational practices that a VIPzone member must 
configure their routers to follow are: 
\begin{enumerate}
\item {\bf Prevent false VERIFIED routes:}
  If the member receives an announcement from a non-member AS, then
  it MUST remove the \ver~community if present.
  This is to prevent an attacker from injecting a hijacked
  route that other VIPzone members prefer.
\item {\bf Drop RPKI-invalid routes:}
  If the member receives an announcement where
  %the prefix owner has registered one or more ROAs that match the prefix, and
  the origin is RPKI-invalid, the member MUST drop the announcement.
  This is to prevent origin hijacks.
\item {\bf Prevent propagation of forged routes:}
  If the member receives an announcement where the AS used by the
  neighbor is not consistent with the AS numbers legitimate for the
  neighbor, the member MUST drop the announcement.
  This is consistent with a Know Your Customer  (KYC)
  requirement, to prevent malicious routes from entering the VIPzone.
\item {\bf Forward VERIFIED routes:}
  If the member receives an announcement from another member with a
  \ver~community tag set, it MUST retain that tag when forwarding the
  route to other members.
  Further, the member MUST retain the \ver~tag when it provides the
  route to non-member neighbors.
  Customers of zone members do not need to understand or act on the \ver~marking, the zone rules allow them to distinguish which routes
  have been \ver~on entry to the zone, and thus are not path hijacks.
\item {\bf Verify routes with one AS in the path from non-member customers:}
  If the member receives an announcement with one AS in the path from
  a non-member customer, it MUST drop the announcement if the route
  contains a prefix that the customer has no authority to announce 
   (it is not RPKI-valid, or is not from a list of prefixes that
  the member has previously established as allowed from their customer).
  If the prefix is RPKI-valid, is registered by the owner in an
  authenticated IRR, or from a list of allowed prefixes, the zone
member AS MUST add a
  \ver~community to the route so that other members know that the
  route is valid.
\item {\bf Forward unverified routes without the VERIFIED tag.}
  If the zone member has not established that the announcement is
  valid (because it has not yet obtained the list of allowed prefixes,
  or because the AS path in the route contains more than one unique ASN and so cannot
  be verified) the member can announce the route to its neighbors but
  MUST NOT add a \ver~community to the route, so that other members do
  not trust the validity of the route.
  To preserve Internet connectivity, zome members must forward unverified 
routes according to normal routing policies.
\item {\bf Export routes to a route collector for auditing.}
  To allow for auditing behavior of trust zone members,
  members must export their routes to a route collector.
\end{enumerate}

\section{Hijack scenarios in a local region}
\label{app:hijacks}
We elaborate on some implications of a local region.
\subsection{Multihoming Transit Scenario}
\label{sec:design:multihoming}
\begin{figure}[t]
\centering
\includegraphics{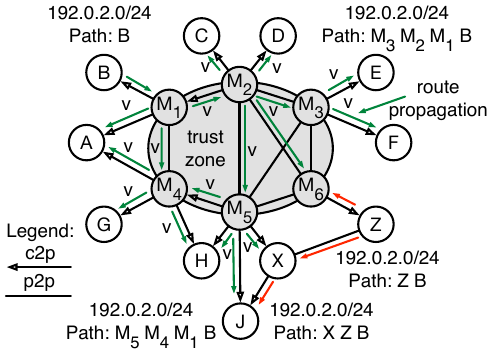}
\caption{The risk of a hijack of J's traffic, in this case destined to
prefix B, is limited to BGP announcements coming from X or Z, but
that assumes it chooses that route rather than a~\ver~route
to B from its zone transit provider M$_{5}$.}
\label{fig:mh3}
\end{figure}

VIPzone members, and non-members exclusively connected to VIPzone
transit providers, will receive an authentic route from the VIPzone if
one is available.  The hijack risk for a VIPzone customer 
increases if it accepts routes from a non-member in its local region.
Figure~\ref{fig:mh3} illustrates the scenario of the residual
risk in a local region from a transit provider that is not in the zone.  Here AS J connects to two transit
providers (M$_{5}$ and X) of which only M$_{5}$ is a zone member. AS X
and AS Z are in the local region of J, since they can originate BGP
announcements that arrive at J without passing through the zone.
If X or Z sends a bogus announcement for a prefix (in this example, B) to J, J might decide
to prefer it over a valid (\ver) route from M$_{5}$. This could happen only if
X or Z are malicious--given the local region of J there are no other ASes
in a position to launch a hijack.

If J did not use X as a transit provider, or preferred the \ver~route
from M$_{5}$, it would prevent this hijack.

\subsection{Use of \ver~ outside the zone}

\begin{comment}
\begin{figure}[t]
\centering
\includegraphics{figures/vipzone-4.pdf}
\caption{The protection that a Routing Trust Zone provides for
  customer routes depends on local regions. 
  If customer J has two transit providers (M$_{5}$ in zone, and K out
  of zone) then other ASes that also have out-of-zone providers (e.g. L)
  may select the unverified route.}
\label{fig:mh1}
\end{figure}
\end{comment}

\begin{figure}[t]
\centering
\includegraphics{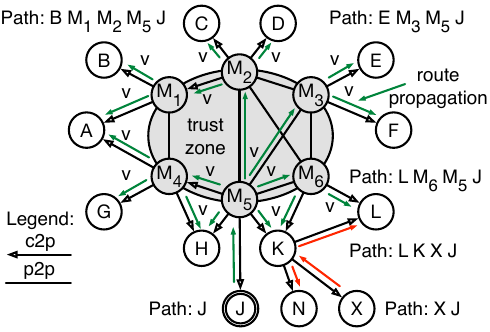}
\caption{An illustration of how protection depends on how 
	customers connect to the Routing Trust Zone.
Customer L receives two routes for J's prefix, a \ver~route via
  M$_{6}$ and an (unverified) hijacked route from X via K.
  If L does not prefer the \ver~route via M$_{6}$ it may select the
  hijacked route because it has the same AS path length.}
  %Our scheme does not provide protection from hijacks 
  %that originate outside the zone in the local region of the 
  %targeted AS (\S\ref{subsec:vipzone-residual-risk}).}
\label{fig:mh2}
\end{figure}

While the VIPzone practices are not required or expected for non-members,
a non-member may choose to configure their routers to remove
\ver~tags from non-member neighbors, and then prefer routes received
from their neighbors who are VIPzone members that are tagged as \ver,
to defend themselves from using a malicious hijacked path towards a
destination (misdirection harm).  VIPzone members must 
retain \ver~tags so that non-members could select these routes.
In Figure~\ref{fig:mh2}, L could have chosen a
\ver~route via M$_{6}$ if L preferred routes received from VIPzone
members that are \ver~ over unverified routes.
 L has an incentive to do so, as otherwise when it
 receives a path hijack by X of equal length to the authentic route
via its VIPzone provider,  it may select the hijacked route and suffer
associated harms.
Note that L has a higher risk of accepting a hijacked route from 
a peer or customer AS as those routes ordinarily have a
higher preference than a provider route. 

\subsection{Peering Interconnection Scenarios}
\label{subsec:vipzone-peering}
In most cases, the analysis for a peering connection is similar to transit
connections.

\subsubsection{Peering with IXP route servers} \label{sec:solution:ixp}

An IXP-operated route server centralizes peering routes from IXP members
and makes these routes available to other IXP members.  If the IXP is a
member of the VIPzone and has configured the route server to verify
routes received from IXP members, then the route server can mark routes
as \ver, and VIPzone members can propagate the \ver~route.  Otherwise,
routes received from a route server are unverified.
(At least one IXP (INEX) has been performing ROV filtering
on its route server since February 2019 \cite{inex-rpki}.)

 \subsubsection{Peering of zone customers outside zone}

If two ASes not in the VIPzone but directly connected
to VIPzone providers peer with each other, they may receive announcements of
routes to each other via the VIPzone that are marked~\ver, and announcements
over the peering connection that are not \ver.  Because ASes not in 
the VIPzone are not expected to use that community value to
assign a preference to an announcement, their routing policy
would be the same as today.  (Note ASes outside the
zone may choose to use this VERIFIED value to prefer routes. 
but they had better know what they are doing, because it may cause unexpected
results, e.g., use of paths via a provider rather than a peer.)

\begin{figure} \vskip 5mm \centering
\includegraphics[width=.4\textwidth]{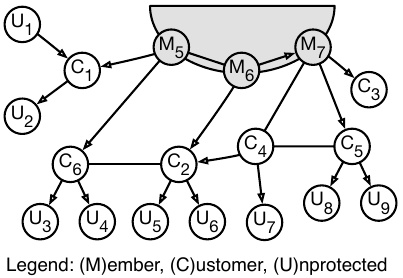}
\caption{{\em {\bf Peering across the VIPzone perimeter.} C$_{2}$ has two
transit providers, only one of which (M$_{6}$) is in the VIPzone. M$_{6}$ will
announce into the VIPzone a verified path to C$_{2}$. C$_{4}$  peers with
M$_{7}$ which is in the VIPzone. C$_{4}$ will announce to M$_{7}$ a route to
C$_{2}$. M$_{7}$ will prefer the \ver~ announcement, and will send traffic to
C$_{4}$  through its provider M$_{6}$ in the zone, not over the peering link to
C$_{4}$. }} \label{fig:transitpeering} \vskip -5mm \end{figure}

\subsubsection{Peering across the VIPzone perimeter}
\label{subsubsec:transit-and-peering}

Peering across the VIPzone perimeter has a straightforward scenario and a
complicated scenario.  Imagine that VIPzone M$_{7}$ in Figure
\ref{fig:transitpeering} peers with non-VIPzone C$_{4}$. In the straightforward
case, M$_{7}$  will apply the same VIPzone rules to peer C$_{4}$ as it does for
customers C$_{3}$ and C$_{5}$, i.e., forward or drop announcements and
mark as \ver~announcements that the peer legitimately originates. 
Customers of that peer C$_{4}$ would not have their routes \ver.  Typical
routing policy is that the AS in the zone would only use these announcements
from peer  C$_{4}$ 
for itself and its customers---it would not forward them on to other
peers or providers.

The complicated peering scenario arises when a customer of that non-zone member also
obtains transit service from an AS in the VIPzone.
Figure~\ref{fig:transitpeering} shows C$_{2}$ with two transit providers, only
one of which is in the zone. The transit provider not in the zone (C$_{4}$)
also peers with an AS in the zone (M$_{7}$). In this case, M$_{7}$ will receive
a \ver~announcement to C$_{2}$ via M$_{6}$, which per the VIPzone rules it must
prefer over the route via the peering link from C$_{4}$, so M$_{7}$ will not
benefit from the peering link for traffic to C$_{2}$, even if it would
normally prefer that peering link.

In \S\ref{sec:considerations}, we plot the number of cases where an AS in the zone would end up using a path through
the zone via a provider, as opposed to the path it would normally prefer (via a peer or customer). Using the current topology of the Internet, we show that for most ASes in a hypothetical VIPzone, the number of such \emph{routing exceptions} is small.

\section{Route leaks}
\label{sec:leaks}

As mentioned in \S\ref{sec:design:vipzone},
a route leak is an event in which an AS inappropriately (i.e.,
violating routing policy) forwards a route it legitimately received.
The consequence is often that large flows of traffic reach this
AS, which is not provisioned to carry them.  A classic
route leak occurs when a multi-homed AS that takes the routes
it receives from one of its transit providers and inadvertently
propagates these routes to its other transit provider.

\begin{figure}
\centering
\includegraphics[width=0.8\linewidth]{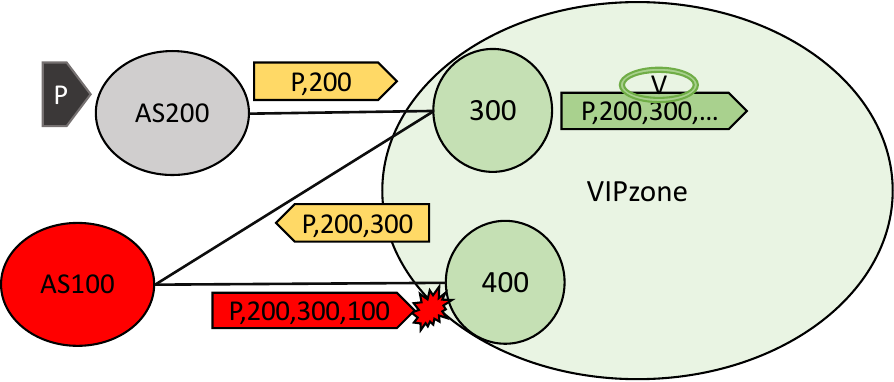}
\caption{AS 100 legitimately receives from its provider AS300 in the zone a route to AS200. If AS400 leaks this route to AS400, the route (which includes multiple AS hops, will not be marked \ver, and since there is a verified announcement for the prefix, it will not be preferred. The leak has no effect.  }
\label{fig:leak}
\end{figure}

In addition to preventing path hijacks of ASes directly attached
to the zone, the VIPzone prevents leaks of announcements of prefixes
belonging to those ASes (Figure~\ref{fig:leak}).
AS 100 might incorrectly announce (leak) the path to AS 200 that it
receives from one transit provider (AS 300) to its other transit provider
(AS 400). Since a \ver~path to AS 200 exists in the zone, AS 400 should
not propagate its unverified route. If it did, ASes in the VIPzone would
never prefer that route, so  customers directly attached to the VIPzone
would not receive that route, and traffic to AS 200 would never flow
from the zone to AS 400.

The VIPzone we have constructed protects against route leaks by ASes
\textit{not} in the VIPzone. If the leak occurs within the zone, the
announcement from AS 300 to AS 100 would be \ver, and when AS100 forwards
(leaked) this announcement to AS 400, AS 400 must remove the \ver marking
first if it propagates the announcement to the zone.

Such potential harms from accidental misconfiguration
suggest an important insight about VIPzone deployment. A natural but unnecessary --- even counterproductive---
objective is to maximize the number of ASes in the VIPzone.
Smaller ASes (certainly stub ASes) will get the benefit of VIPzone
from being a \textit{customer} of a VIPzone member.  Actually joining
will require that the joining AS correctly implement
a range of operational practices,
which for smaller ASes with less sophisticated staff may be difficult.
Getting these practices wrong may result in malformed announcements in
the zone, which will lead to the revocation of their VIPzone status.
We consider it preferable that only operators
with sufficient technical abilities attempt to
join the VIPzone. Other requirements (such
as maintaining correct contact information, registering their own
prefixes in a public database, implementing anti-spoofing filters)
make sense for an AS of any size, and a MANRS-like initiative may want
to define two tiers of ISP membership to accommodate different
likely capabilities.

\end{appendix}

\end{document}